\documentclass[apj]{emulateapj}
\usepackage{psfig}

\def\axp{\hbox{AX~J1844.8$-$0256}}
\def\taxp{\hbox{XTE~J1810$-$197}}

\newcommand\xte{{\it RXTE\/}}

\newcommand\einstein{{\it Einstein}}

\newcommand\asca{{\it ASCA\/}}

\newcommand\rosat{{\it ROSAT\/}}
\newcommand\chandra{{\it Chandra}}

\newcommand\xmm{{\it XMM-Newton}}

\shorttitle{Anomalous X-ray Pulsar XTE J1810--197}
\shortauthors{Halpern and Gotthelf}

\slugcomment{Received 2004 April 30; accepted 2004 September 22}
\begin{document}

\title{Fading of the Transient Anomalous X-ray Pulsar\\ XTE J1810--197}
\author{J. P. Halpern and E. V. Gotthelf}
\affil{Columbia Astrophysics Laboratory, Columbia University, 550 West
120$^{th}$ Street,\\ New York, NY 10027-6601; jules@astro.columbia.edu}

\begin{abstract}

Three observations of the 5.54~s Transient Anomalous X-ray Pulsar \taxp\
obtained over 6 months with the {\it Newton X-Ray Multi-Mirror Mission}
(\xmm) are used to study its spectrum and pulsed light curve
as the source fades from outburst.   The decay is
consistent with an exponential of time constant $\approx 300$ days,
but not a power law as predicted in some models of sudden
deep crustal heating events. All spectra are well fitted by a
blackbody plus a steep power law, a problematic model that is commonly
fitted to anomalous X-ray pulsars (AXPs).  A two-temperature
blackbody fit is also acceptable, and better motivated physically
in view of the faint optical/IR fluxes, the X-ray pulse shapes
that weakly depend on energy in \taxp, and the inferred emitting areas
that are less than or equal to the surface area of a neutron star.
The fitted temperatures remained the same
while the flux declined by 46\%, which can be interpreted
as a decrease in area of the emitting regions.
The pulsar continues to spin down, albeit at a
reduced rate of $(5.1 \pm 1.6)\times10^{-12}$ s~s$^{-1}$. The inferred
characteristic age $\tau_{\rm c} \equiv P/2\dot P \approx 17,000$~yr,
magnetic field strength $B_{\rm s} \approx 1.7\times 10^{14}$~G,
and outburst properties are consistent with both the outburst and 
quiescent X-ray luminosities
being powered by magnetic field decay, i.e., \taxp\ is a magnetar.

\end{abstract}

\section{Introduction}

\begin{figure}
\vskip 0.1in
\plotone{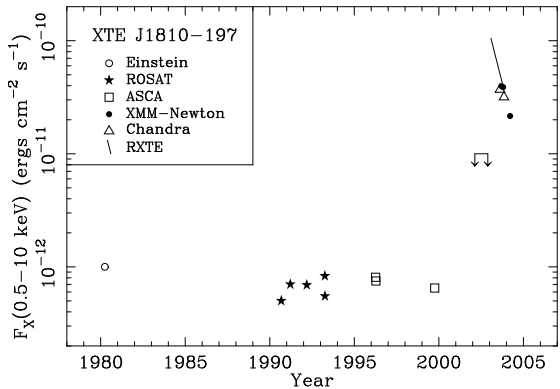}
\caption{The $0.5-10$ keV flux history of \taxp\
spanning 24 years.  The historical
data points are from Paper~I.
The {\it solid line} is an exponential fit 
with time constant 269 days to the \xte\ monitoring
from \citet{ibr04}.  Upper limits indicated in 2002 are also
from several months of monitoring by \xte, which constrains the onset
of the outburst to $<2$ months before its initial detection
in 2003 January. \label{history}}
\end{figure}

The bright 5.54~s X-ray pulsar \taxp\ was discovered serendipitously by \citet{ibr04}
using the {\it Rossi X-ray Timing Explorer} (\xte), and was
localized in two target-of-opportunity (TOO) observations 
with the {\it Chandra X-ray Observatory}, reported by
\citet[hereafter Paper~I]{got04} and \citet{isr04}.
The source had became active sometime between 2002 November 17
and 2003 January 23.  The maximum (2--10~keV) flux observed by \xte\ was
$\approx 6 \times 10^{-11}$ 
ergs~cm$^{-2}$~s$^{-1}$, when it was already declining
with an exponential time constant of $269\pm25$~days \citep{ibr04}.
Of unique value are all of the archival \einstein, \rosat, and \asca\
detections at the location of \taxp,
which indicate a long-lived quiescent baseline flux of
$F_x (0.5-10\,{\rm keV}) \approx 7 \times 10^{-13}$ ergs~cm$^{-2}$~s$^{-1}$
for at least 13~years and possibly for 23~years prior to the
onset of the active state (Paper~I and Figure~\ref{history}).
Timing of the highly modulated signal over the first 9 months
of the outburst indicated
rapid spin-down with characteristic age $\tau_{\rm c}
\equiv P/2\dot P \leq 7600$~yr, surface magnetic field $B_{\rm s}
\approx 2.6\times 10^{14}$~G, and spin-down power $\dot E \approx 4
\times 10^{33}$ erg~s$^{-1}$ \citep{ibr04}.  Deep IR observations \citep{isr04}
detected a faint candidate of $K_s = 20.8$,
similar to ones associated with other AXPs,
within the final $0.\!^{\prime\prime}6$ radius \chandra\ error circle of \taxp.
Variability of the IR source confirmed its identification with \taxp\
\citep{rea04a,rea04b}.

The magnetar model \citep{dun92} seems poised to address
observational properties of several categories of
single, young neutron stars (NSs) that are not powered
by rotation.  These include the AXPs, Soft Gamma-ray
Repeaters (SGRs), and Central Compact Objects (CCOs)
within supernova remnants that lack detected periods
but whose spectra resemble those of AXPs and SGRs.
Such unification is supported by the discovery of
transient behavior in AXP-like objects.
The first suggested TAXP was \axp\ \citep{got98}.
The timing and spectral properties of that 7-s X-ray pulsar
are consistent with an AXP, while its flux was found to change
by orders-of-magnitude from an ``active'' pulsar state to a faint
``quiescent'' state since its discovery.  However, the lack of a $\dot P$
measurement for \axp, which was observed serendipitously only once,
has left its status as an AXP unconfirmed.
With the discovery of the {\it bona fide} TAXP \taxp,
consideration of possible related observations
such as the cooling trend in the post-burst fluxes
of SGR 1627--41 \citep{kou03}, SGR 1900+14 \citep{len03},
and AXP 1E~2259+586 \citep{woo04},
and the variable CCO in the young SNR RCW~103 \citep*{got97,got99a}, 
motivates development of a unified magnetar model
for the entire AXP/SGR/CCO family.

Although TAXPs are relatively rare, their short active
duty cycle suggests the existence of a significant number of
as-yet unrecognized, although not necessarily undetected, young NSs.
\taxp\ provides a new window into this population, with a
well documented quiescent history and detailed observations
during its active, pulsed phase.  Monitoring of the fading
of \taxp\ takes advantage of a possibly unique opportunity
to study evolving spectral and timing behavior of a TAXP,
which reveals the time-dependence of the thermal and non-thermal
emission mechanisms to which magnetars convert their energy.
In particular, the physical interpretation of the two-component X-ray spectrum
that seems characteristic of most AXPs can be investigated in more
ways than is possible for a quiescent source.  We are beginning
such a study with the \xmm\ observations of \taxp\
listed in Table~\ref{log}.

\begin{deluxetable*}{lccc}
\tablewidth{0pt}
\tablecaption{\xmm\ Observations of \taxp\ \label{log}}
\tablehead{
                     & \colhead{EPIC pn}	  & \colhead{EPIC pn}  & \colhead{EPIC pn} \\
\colhead{Parameter}  & \colhead{2003 Sep 8} & \colhead{2003 Oct 12} & \colhead{2004 Mar 11}
}
\startdata
Off-axis angle (arcmin)                   &      0.0           &     8.8           &      0.0          \\
Exposure time (ks)                        &     11.5           &     6.9           &     17.0          \\
Live Time (ks)                            &      8.1           &     6.2           &     15.8          \\
Count Rate (s$^{-1}$)\tablenotemark{a}    &     10.6           &     4.8           &      5.8          \\
Epoch (MJD/TDB)\tablenotemark{b}          &  52890.5642044     &   52924.0000320   &  53075.4999960    \\
Period (s)\tablenotemark{c}               &      5.539344(19)\tablenotemark{c}     &      5.53943(8)
                                          &      5.539425(16)\tablenotemark{c}                         \\
\enddata
\tablenotetext{a}{\footnotesize Background subtracted EPIC pn count rate corrected for detector dead-time.}
\tablenotetext{b}{\footnotesize Epoch of phase zero in Figure~\ref{pulse}.} 
\tablenotetext{c}{\footnotesize Includes EPIC MOS data.  95\% confidence uncertainty in parentheses.}
\end{deluxetable*}

\section{\xmm\ Observations and Results}

The first \xmm\ Target of Opportunity (TOO)
observation of \taxp\ was obtained on
2003 September 8 and described in Paper~I.
A second \xmm\ TOO observation of \taxp\ was performed 
on 2004 March 11.  We use the
data obtained with the European Photon Imaging Camera
(EPIC, \citealt{tur03}).  EPIC consists of three CCD cameras,
the EPIC pn and the two EPIC MOS imagers,
which are sensitive to X-rays in $0.1-10$~keV nominal range.
The observing modes were different for the two observations.
In 2003 September, one of the EPIC MOS cameras was operated
``small window'' mode, for which the field of view (FOV)
of the central CCD was reduced to $1\farcm8 \times 1\farcm8$ and
read out in a time of 0.3~s.
The second EPIC MOS camera was operated
in ``full frame'' mode, which is read out in 2.7~s integrations
over the $30^{\prime}$ FOV.
The EPIC pn was operated in ``small window'' mode, which provides 6~ms
time resolution over a $4\farcm3 \times 4\farcm3$ FOV.
In 2004 March, both of the EPIC MOS cameras were
operated in ``small window'' mode, while the
EPIC pn was operated in ``large window'' mode, which provides
48~ms time resolution over a $13\farcm5 \times 26^{\prime}$ FOV.
Thus, all data sets except the ``full frame'' MOS in 2003
September were of high enough time resolution to study the
pulse profiles of \taxp.  In addition, \taxp\ was detected
serendipitously in a short observation, as yet unpublished,
from the \xmm\ Galactic Plane Survey \citep{han04} on 2003 October 12.
We include results from this survey observation for completeness,
including timing information from the pn camera in ``full frame''
mode with 73.4~ms time resolution.	

In order to make the cleanest possible comparison among the
\xmm\ observations, we extracted all calibrated photon
event data files using the latest version of the \xmm\
Science Analysis System [release version SAS 6.0.0 (20040309\_1146)],
by applying the processing chains to each set of Observation Data Files (ODF).
Screened photon event lists were generated by applying the standard \xmm\
good-time filter criteria and selecting CCD PATTERN
$\le 4$ for our EPIC pn spectral analysis.  A total of
11.5~ks, 6.9~ks, and 17.0~ks of good EPIC pn exposure time
were acquired during the three observing epochs,
which translates to an effective live time of
8.1~ks, 6.2~ks, and 15.8~ks.  Resulting background subtracted count rates
were 10.6~s$^{-1}$, 4.8~s$^{-1}$, and 5.8~s$^{-1}$, where the second one
is affected by vignetting at the off-axis angle of $8.\!^{\prime}8$.
The observations were mostly uncontaminated by flare events.  
Photon arrival times were converted to
the solar system barycenter using the \chandra\ derived coordinates of
the source given in Paper~I,
(J2000) $18^{\rm h}09^{\rm m}51.\!^{\rm s}08$, $-19^{\circ}43^{\prime}51.\!^{\prime\prime}7$.
We manually corrected a jump of $+1$~s in the EPIC pn photon arrival times
that affected part of the 2004 March observation.   Timing anomalies
are frequent in the pn data; evidently, they have not been completely
eliminated in SAS 6.0.0.

\subsection {Spectral Modeling}

In Table~\ref{spectra}, we summarize the results
of spectral analysis of previous and new observations.
Although of relatively poor quality, the quiescent \rosat\ spectrum
of \taxp\ can be reasonably well fitted with a blackbody of $kT = 0.18\pm 0.02$~keV
covering $1.2\times10^{13}$~cm$^2$, or 2/3 the area of a 12~km radius NS at
$d = 5$~kpc.  In contrast, the X-ray spectra in outburst, obtained from the
\xmm\ TOO observations, required two components, which is typical of
AXPs \citep[e.g.,][]{mar01}.
We fitted \xmm\ data from the EPIC pn CCD only. The fast
read out of this instrument ensures that its spectrum is not affected
by photon pile-up.  Source spectra were accumulated
in a $45^{\prime\prime}$ radius aperture which encloses $\geq 95\%$ of the
encircled energy.  Background was taken from a circle of the same size
displaced $2\farcm3$ along the readout direction.
The spectra were grouped into bins containing a minimum of 400
counts (including background) and fitted using the XSPEC package. 
The ``wabs'' photoelectric absorption cross sections of \citet{mor83},
not including Thomson scattering, were used with the \citet{and82} abundances.
For each model, the column density was treated as a single parameter
in a linked fit to all three epochs, after verifying that
independent fits yielded consistent values of $N_{\rm H}$.
The other fitted parameters are also consistent with their
values obtained in fits with independent $N_{\rm H}$.

\begin{deluxetable*}{lcccc}
\tablewidth{0pt}
\tablecaption{Spectral Fits and Fluxes\label{spectra}}
\tablehead{
\colhead{Model}   & \colhead {\rosat /PSPC} & \colhead{}	 & \colhead{\xmm /EPIC pn} \\
\colhead{Parameter}  & \colhead{1992 Mar 7} & \colhead{2003 Sep 8} & \colhead{2003 Oct 12} & \colhead{2004 Mar 11}
}
\startdata
Blackbody \\
\tableline
$N_{\rm H}$ ($10^{22}$ cm$^{-2}$)		& $0.63$ (fixed)	& . . .	& . . .	& . . .  \\
$kT_{\rm BB}$ (keV)				& $0.18\pm0.02$		& . . .	& . . .	& . . .  \\
BB Area (cm$^2$)				& $1.2\times10^{13}$	& . . .	& . . .	& . . .  \\
Flux (ergs cm$^{-2}$ s$^{-1}$)\tablenotemark{a} & $6.9 \times10^{-13}$ 	& . . .	& . . .	& . . .  \\
$L_{\rm BB}$(bol) (ergs s$^{-1}$)               & $1.3 \times10^{34}$ 	& . . .	& . . .	& . . .  \\
$\chi^2_{\nu}$(dof)				& 1.1(13)		& . . .	& . . .	& . . .  \\
\tableline
Power-law + Blackbody \\
\tableline
$N_{\rm H}$ ($10^{22}$ cm$^{-2}$)\tablenotemark{b}&. . .& $1.02 \pm 0.05$	& $1.02 \pm 0.05$ & $1.02 \pm 0.05$  \\
$\Gamma$					&. . .	& $3.87 \pm 0.25$	& $3.51 \pm 0.15$ & $3.75 \pm 0.20$  \\
$kT_{\rm BB}$ (keV)				&. . .	& $0.67 \pm 0.01$ 	& $0.68 \pm 0.01$ & $0.68 \pm 0.01$  \\
BB Area (cm$^2$)				&. . .	& $6.2\times10^{11}$	& $5.2\times10^{11}$	& $2.7\times10^{11}$ \\
PL Flux (ergs cm$^{-2}$ s$^{-1}$)\tablenotemark{a}&. . .& $1.17\times10^{-11}$  & $1.42\times10^{-11}$  & $8.47\times10^{-12}$\\
$L_{\rm PL}$ (ergs s$^{-1}$)\tablenotemark{c}   &  . . .& $2.7\times10^{35}$    & $3.6\times10^{35}$    &  $1.8\times10^{35}$  \\
BB Flux (ergs cm$^{-2}$ s$^{-1}$)\tablenotemark{a}
						& . . .	& $2.80\times10^{-11}$  & $2.47\times10^{-11}$  &
$1.31\times10^{-11}$ \\
$L_{\rm BB}$(bol) (ergs s$^{-1}$)               & . . .	& $1.3\times10^{35}$    & $1.1\times10^{35}$    & $6.0\times10^{34}$  \\
Total Flux (ergs cm$^{-2}$ s$^{-1}$)\tablenotemark{a}
						&. . .	& $3.97\times10^{-11}$	& $3.89\times10^{-11}$ 	& $2.16\times10^{-11}$ \\
$\chi^2_{\nu}$(dof)\tablenotemark{b}		&. . .	& 1.077(187) & 1.077(84)	& 1.077(194) \\
\tableline
Double Blackbody \\
\tableline
$N_{\rm H}$ ($10^{22}$ cm$^{-2}$)\tablenotemark{b}&. . .& $0.65 \pm 0.04$	& $0.65 \pm 0.04$ & $0.65 \pm 0.04$  \\
$kT_1$ (keV)					&. . .	& $0.26 \pm 0.02$	& $0.29 \pm 0.04$ & $0.27 \pm 0.02$  \\
$kT_2$ (keV)					&. . .	& $0.68 \pm 0.02$	& $0.71 \pm 0.03$ & $0.70 \pm 0.01$  \\
BB1 Area (cm$^2$)				&. . .	& $1.1\times10^{13}$	& $6.6\times10^{12}$ & $6.8\times10^{12}$\\
BB2 Area (cm$^2$) 				&. . .	& $6.4\times10^{11}$	& $5.1\times10^{11}$ & $2.9\times10^{11}$\\
BB1 Flux (ergs cm$^{-2}$ s$^{-1}$)\tablenotemark{a}
						& . . .	& $4.24\times10^{-12}$	& $5.37\times10^{-12}$	& $3.52\times10^{-12}$ \\
BB2 Flux (ergs cm$^{-2}$ s$^{-1}$)\tablenotemark{a}
						& . . .	& $3.51\times10^{-11}$  & $3.03\times10^{-11}$  & $1.78\times10^{-11}$ \\
$L_{\rm BB1}$(bol) (ergs s$^{-1}$)              & . . . & $5.2\times10^{34}$ 	& $5.1\times10^{34}$ 	&  $3.9\times10^{34}$ \\
$L_{\rm BB2}$(bol) (ergs s$^{-1}$)              & . . . & $1.4\times10^{35}$    & $1.3\times10^{35}$ 	&  $7.2\times10^{34}$ \\
Total Flux (ergs cm$^{-2}$ s$^{-1}$)\tablenotemark{a}
						&. . .	& $3.93\times10^{-11}$	& $3.84\times10^{-11}$	& $2.13\times10^{-11}$\\
$\chi^2_{\nu}$(dof)\tablenotemark{b}		&. . .	& 1.076(187) & 1.076(84) & 1.076(194)  \\ 
\enddata
\tablenotetext{a}{Absorbed flux in the 0.5--10 keV band.}
\tablenotetext{b}{Parameter derived from a linked fit to all epochs.}
\tablenotetext{c}{Unabsorbed luminosity in the 0.5--10 keV band assuming $d = 5$~kpc.}
\tablecomments{Uncertainties are 90\% confidence for two interesting parameters.} 
\end{deluxetable*}

\begin{figure*}
\epsscale{1.3}
\vskip -0.7in
\centerline{
\plotone{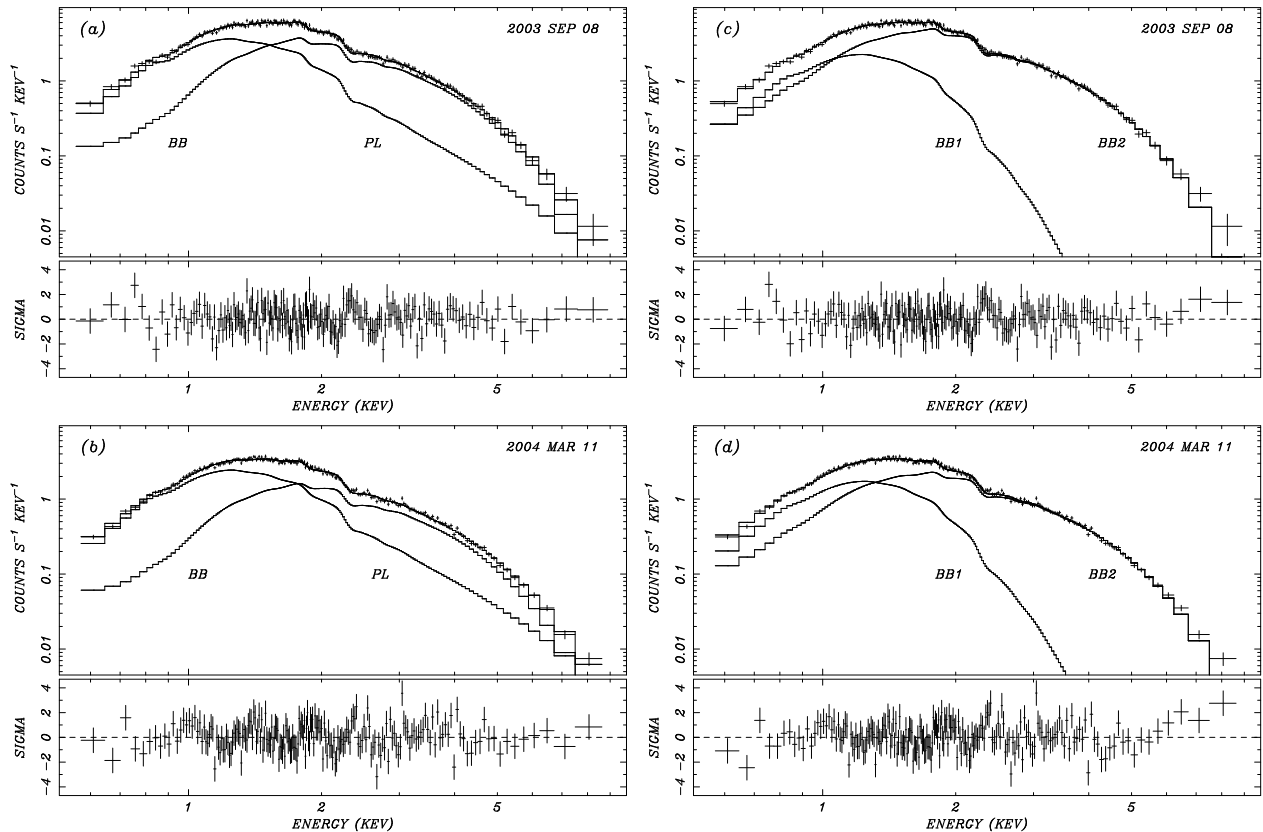}
}
\vskip -4.8in
\caption{\xmm\ EPIC pn spectra of \taxp\ from the earliest and latest epochs,
fitted with a power-law plus blackbody model or a double
blackbody model as described in the text and in Table~\ref{spectra}.
Also shown are the residuals from the best-fit models.
({\it a\/}) Power-law plus blackbody fit in 2003 September.
({\it b\/}) Double blackbody fit in 2003 September.
({\it c\/}) Power-law plus blackbody fit in 2004 March.
({\it d\/}) Double blackbody fit in 2004 March.
\label{specfig}}
\end{figure*}

The \xmm\ spectra are equally well fitted by a power-law plus
blackbody model, as commonly quoted for AXPs,
or by a two-temperature blackbody model that
we will argue is more physically motivated.
Sample fits are shown in Figure~\ref{specfig}.
But the ambiguity about which X-ray spectral model is appropriate
affects our estimate of the distance to the source by contributing
systematic uncertainty to the fitted column density.
The X-ray measured $N_{\rm H} = (1.02 \pm 0.05) \times 10^{22}$~cm$^{-2}$ in
the case of the power-law plus blackbody model is significantly
different from the value of $(0.65 \pm 0.04) \times 10^{22}$~cm$^{-2}$
in the double blackbody model (see Table~\ref{spectra}).
We argued in Paper~I that,
in either case, \taxp\ is likely to be closer than the
neighboring supernova remnant G11.2--0.3, which has an
H~I absorption kinematic distance of 5~kpc \citep{gre88}
and an X-ray measured column density of $\approx 1.4
\times 10^{22}$~cm$^{-2}$ \citep{vas96}.
With the distance somewhat uncertain,
we parameterize derived quantities in terms of $d_5$,
an upper limit on the distance, in units of 5~kpc.

Considering first the 2003 September observation,
the best fit to the power-law plus blackbody model 
has photon index $\Gamma = 3.87\pm0.25$ and $kT_{\rm BB} = 0.67\pm 0.01$~keV,
with a fit statistic of
$\chi^2_{\nu} = 1.077$ for 187 degrees of freedom. 
The flux for each of these components is given in Table~\ref{spectra}.
The blackbody flux is equivalent to a spherical area of
$6.2 \times 10^{11}\,d_5^2$~cm$^2$,
or $\approx 3\%$ of the surface area of a NS.
It is noteworthy that the 2004 March spectrum had nearly identical
parameters, with $\Gamma = 3.75\pm0.20$ and $kT_{\rm BB} = 0.68\pm 0.01$~keV,
while the fluxes of the power law and blackbody
declined by 28\% and 53\%, respectively.

A two-temperature blackbody model yielded a fit of
essentially the same quality, with a fit statistic of
$\chi^2_{\nu} = 1.076$ for 187 degrees of freedom.
The 2003 September spectrum has $kT_1 = 0.26\pm 0.02$~keV
and $kT_2 = 0.68\pm 0.02$~keV associated with blackbody areas
of $1.1 \times 10^{13}\,d_5^2$~cm$^2$ and 
$6.4 \times 10^{11}\,d_5^2$~cm$^2$ for the cooler
and hotter temperatures, respectively, which are $\approx 60\%$
and $\approx 4\%$ of the NS surface.
Thus, the power-law component of the previous model is replaced with a
cooler blackbody, while the hotter thermal component retains
the same parameters in both models.
In the double blackbody model, the cooler component
covers a large fraction of the NS surface, but at a
temperature significantly higher than the quiescent \rosat\
measured value.  In this parameterization, the cooler and hotter
blackbody luminosities declined by 25\% and 50\%, respectively,
between 2003 September and 2004 March.
Since their temperatures remained the same to within 3\%
($kT_1 = 0.27\pm 0.02$~keV and $kT_2 = 0.70\pm 0.01$~keV in 2004 March), 
their decline in luminosity is attributable to a decrease in area.

In order to extract further information about the distribution of
emission over the stellar surface,
we analyzed phase-resolved spectra by fitting the two-temperature
blackbody model to the 2003 September data in 10 phase bins.  In
particular, we kept the temperatures and column density 
fixed at the values listed in
Table~\ref{spectra} while fitting for the normalization (intensity)
in each phase bin.  This is equivalent to determining the
relative projected area of emission as a function of rotation phase.
This procedure yields acceptable fits in all phase bins, with
reduced $\chi^2$ values ranging from 0.73 to 1.21.  In nine out
of 10 bins, the reduced $\chi^2$ is less than 1.08.
Since these fits are statistically acceptable as a set,
no significant test can be made for variations in
additional parameters such as the temperatures or column density.

\begin{figure}
\vskip 0.15in
\centerline{
\psfig{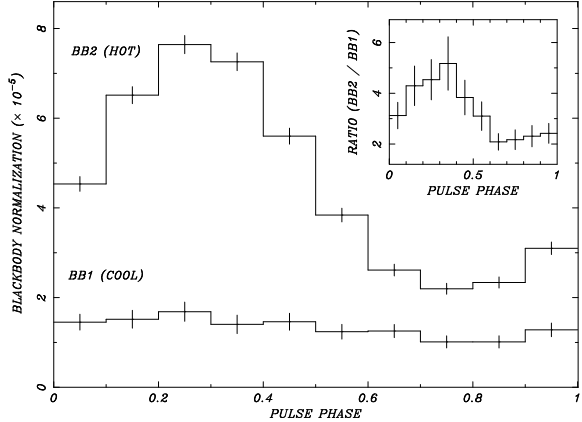}
}
\caption{Comparison of the two-component blackbody intensities as
a function of rotation phase in the 2003 September \xmm\ EPIC pn spectrum.
The temperatures $kT_1$ and $kT_2$ are held fixed at the values listed
in Table~\ref{specfig}, while the normalization constants are fitted.
{\it Inset\/}: The ratio of blackbody normalization constants as a function
of rotation phase.
\label{phase}
}
\vskip -0.05in
\end{figure}

The results of the phase-resolved spectral fitting
are shown in Figure~\ref{phase}.  As can be expected, the
intensity of the hot component varies by a large factor,
$\approx 3.5$, while the larger cool component varies by only a
factor of 1.7.  Both components peak at the same phase.
This is consistent with the energy-dependent light
curves (Fig.~\ref{pulse}), in which the pulsed fraction increases
with increasing energy.  The relative amplitudes and phases of the
two components are therefore consistent with the picture of a small,
hot region surrounded by a cooler, concentric annulus that occupies
of order half the surface area of the star.  Neither component goes to
zero at any rotation phase.  In particular, if the hot component were
completely eclipsed, the spectral decomposition in Figure~\ref{specfig}
indicates that the light curves at $E > 3$~keV should go to zero,
which clearly they do not.

While the overall fits of the two models are of equal quality,
the double blackbody model does leave significant residuals above
6~keV, which may be indicative of an additional harder component,
especially in the 2004 March spectrum.  We do not have
enough high-energy information to model this excess quantitatively,
but it may represent nonthermal magnetospheric
emission similar to the hard X-ray component recently reported
from the AXPs 1E~1841--045 \citep*{bas04,kui04,mol04},
1RXS J170849.0--400910 \citep{rev04},
and 4U 0142+614 \citep{den04} in \xte\ and {\it INTEGRAL\/} data.

\subsection {Pulsed Light Curves and Period Derivative}

\begin{figure}
\vskip -0.3in
\epsscale{1.3}
\centerline{
\plotone{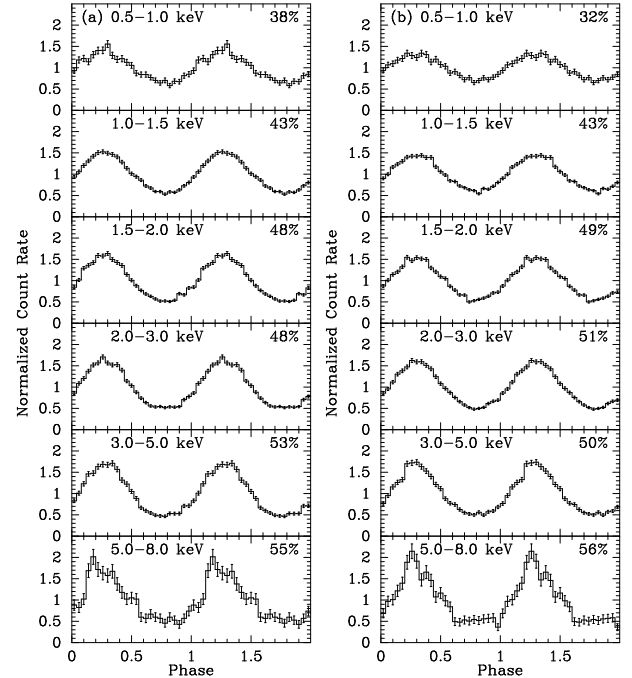}
}
\vskip -0.8in
\caption{Energy-dependent pulse profiles of \taxp.
({\it a\/}) EPIC pn on 2003 September 8.
({\it b\/}) EPIC pn on 2004 March 11.
The epochs of phase zero are given in Table~\ref{log}.
Pulsed fractions, defined as the fraction of the counts above the minimum
in the light curve, are indicated in each panel.
Background has been subtracted.  Pulse profiles from 2003 October 12
(not shown) are similar to these.
\label{pulse}
}
\vskip -0.2in
\end{figure}

The quiescent \rosat\ source was not modulated, with a pulsed fraction
upper limit of 24\% (Paper~I).
Since the first detection of pulsed emission in early
2003, the pulsed fraction remained fairly constant at $\approx 54\%$
as seen in the two \chandra\ observations on 2003 August 27 and November 1
(Paper~I).
The folded light curves from the 2003 September 8 \xmm\ observation
(Fig.~\ref{pulse}) show that the pulse peak is somewhat narrower than a sinusoid,
and the pulsed fractions increase smoothly with energy from 
38\% at less than 1~keV to $\approx 55\%$ above 5~keV.  
As discussed in \S 2.1, the fact that the high-energy
light curves do not dip to zero implies that the hot blackbody
component is never completely out of view.  In 2003 October and
2004 March, the \xmm\ light curves have not changed much.  The pulse
shape and pulsed fractions are largely unchanged.  The marginal appearance
of narrow components at the peak of the pulse profile in the highest
energy bin may be additional evidence of a hard nonthermal contribution.

The measured barycentric pulse period was $5.539344 \pm 0.000019$~s
on 2003 September~8, and $5.539425 \pm 0.000016$~s on 2004 March~11 
(Table~\ref{log}), from the combined EPIC pn and MOS data.
The measurement from the shorter pn observation of 2003 October~12
has a much larger error bar, but is consistent with the other
values.  The errors are the 95\% confidence level determined from the
$Z_1^2$ test.  The two precise period measurements can be compared
to yield $\dot P = (5.1 \pm 1.6) \times 10^{-12}$ s~s$^{-1}$
over the 6~month interval, which implies a characteristic age $\tau_{\rm c}
\approx 17,000$~yr, surface magnetic field $B_{\rm s}
\approx 1.7\times 10^{14}$~G, and spin-down power $\dot E \approx 1.2
\times 10^{33}$ erg~s$^{-1}$, comparable to the earlier
values measured by \citet{ibr04}.  However, it is evident that
$\dot P$ is not stable, having varied by at least a factor
of 2 on timescales of months.  \citet{ibr04} fitted values
of $\dot P$ in the range $(1.1-2.2) \times 10^{-11}$~s~s$^{-1}$
in the first 9 months of the outburst.  Increases in $\dot P$ of
order unity have been seen in the AXPs 1E~2259+586 \citep*{iwa92,kas03}
and 1RXS J170849.0--400910 \citep{dal03} following their glitches.
Even larger changes in $\dot P$ are seen in the 
SGRs 1900+14 \citep{kou99,woo99,woo03} and 1806--20 \citep{woo00,woo02},
although these were apparently not associated with episodes of bursting.
It is possible that \taxp\ experienced a glitch and an increase in
$\dot P$ at the time of its outburst, and that $\dot P$ is now
relaxing to its long-term value.  Alternatively, the spin-down
could have been enhanced in the early stages of the outburst
by a particle wind that is now declining.  Sufficiently
energetic Alfv\'en waves drive winds that can increase the
magnetic field strength at the light cylinder over the dipole value,
thereby increasing the spin-down torque.
\citep{tho98}.

We note here that two discrepant period measurements have been
reported from independent analyses of the 2004 March observation.
\citet{rea04a} gave $P = 5.53974 \pm 0.00005$~s,
and later \citet{rea04b} gave $P = 5.539917 \pm 0.000005$~s.
Both of these are evidently affected by the error in the EPIC pn timing
that we discovered in the data (\S 2).  Consequently, the much larger
$\dot P$ reported by \citet{rea04b} is not valid.  Our correct
period measurement in Table~\ref{log}, $P = 5.539425 \pm 0.000016$~s,
is confirmed by establishing consistency, both in period and in phase,
among the three EPIC instruments' data analyzed separately.

\subsection {Long-term Flux Decay}

\citet{ibr04} fitted the decline of the flux from \taxp\ as observed
by \xte\ with either an exponential of time constant $269\pm25$~days,
or a power law of $t^{-\beta}$ where $0.45 \leq \beta \leq 0.73$.
The allowed range of $\beta$ comes from the uncertainty in the initial
epoch of the outburst, 2003 November 17 $< t_0 <$ 2004 January 23.
The new \xmm\ observations extend 
the baseline by an additional 5.5~months, although
it is difficult to make a precise comparison
between \xte\ and \xmm\ fluxes because of the very soft spectrum
of the source, and the considerable background and source confusion
that plagues \xte\ in this region of the Galactic plane.
Nevertheless, the first
\xmm\ observation is consistent in flux with the contemporaneous
\xte\ monitoring (Fig.~\ref{history} and \citealp{ibr04}), and the
September-March \xmm\ points alone fit an exponential time constant of 300~days,
marginally consistent with the \xte\ determined decay.  However, if
we fit the \xmm\ points with a power-law decay, then the allowed
exponent range is $1.03 \leq \beta \leq 1.26$, quite a bit steeper
than the \xte\ fitted rate.  Continued monitoring is thus able to
discriminate among exponential, power law, or more complicated decays.
It already seems that an exponential is a better fit than a power law.

The total energy emitted in the outburst can be estimated by 
integrating the fitted exponential decay, i.e., multiplying the
peak luminosity by the decay constant.
Assuming a peak bolometric luminosity of $6 \times 10^{35}\,d_5^2$
ergs~s$^{-1}$ at the onset of the \xte\ observered outburst 
\citep{ibr04} and a time constant of 300 days yields a fluence
of $1.5 \times 10^{43}\,d_5^2$~ergs.

\section{Discussion}

\subsection{Problems with a Power-Law X-ray Spectrum}

A two-component X-ray spectral fit is commonly required for AXPs,
but its physical significance is poorly understood.  While it is
conventional to fit a power-law plus blackbody model, the
interpretation of the resulting steep power law,
e.g., $\Gamma \approx 3.8$ in the case of \taxp,
is not clear.  When employing power laws as a means
of fitting excess high-energy emission, problems arising
from the dominance of such power laws at low energy are
usually ignored.  Here, we point out that such a steep power
law cannot be connected to the faint IR fluxes measured by 
\citet{isr04} without assuming severe and perhaps
unphysical cut-off mechanisms.

\begin{figure*}
\vskip 0.1in
\epsscale{1.1}
\plottwo{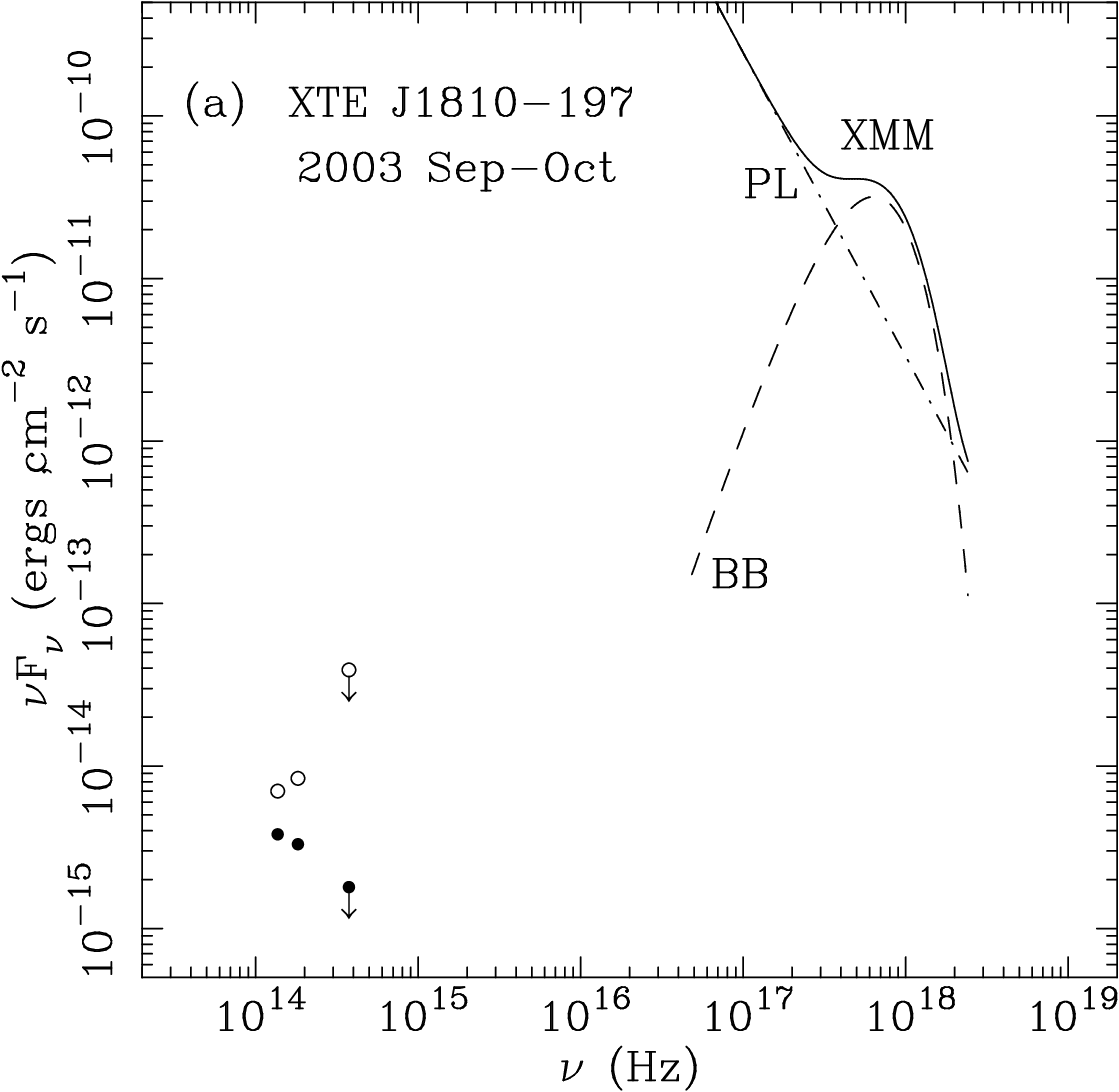}{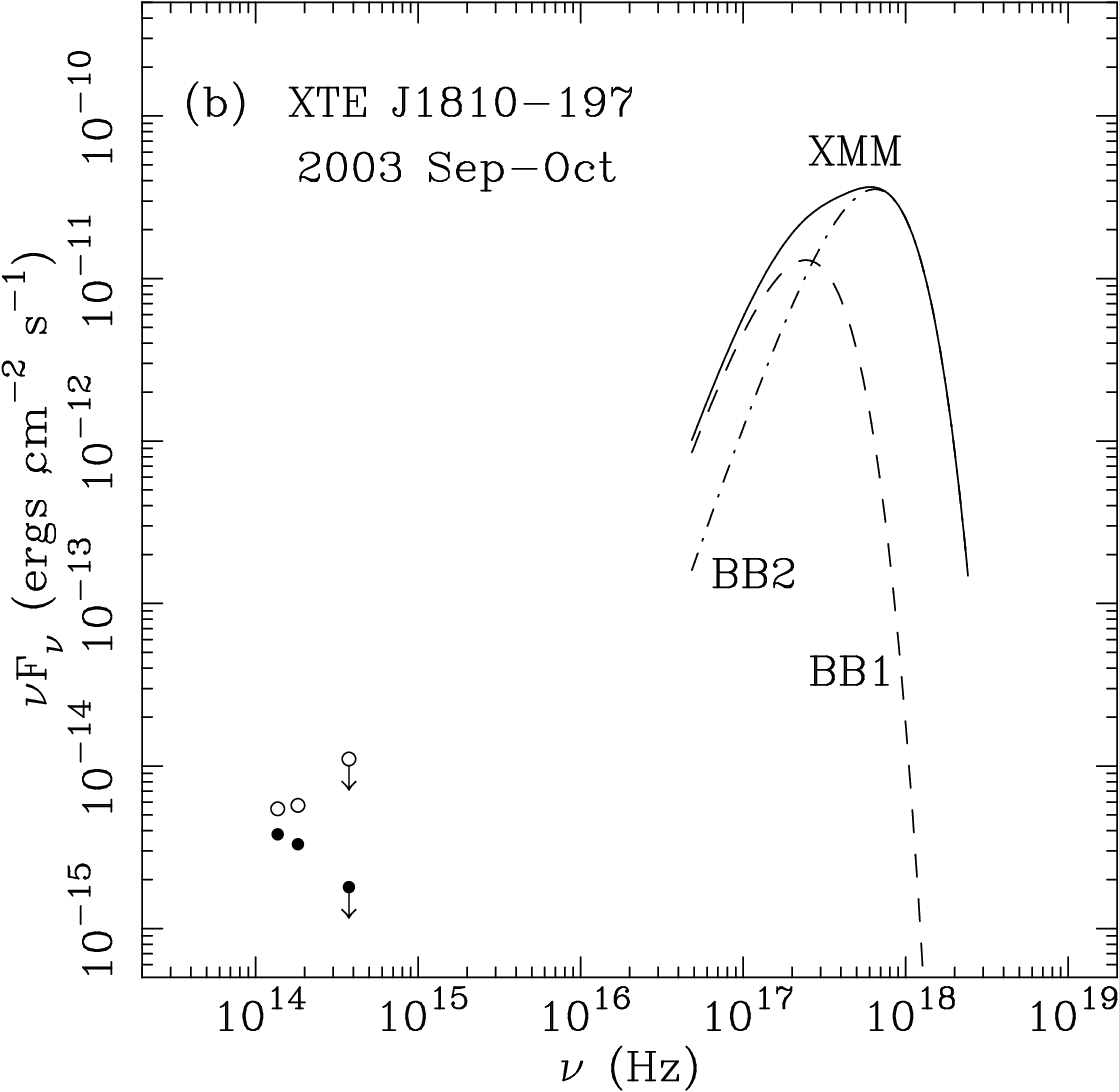}
\caption{Broad-band spectrum of \taxp\ as derived
from fits to the \xmm\ EPIC pn data of 2003 September, and contemporaneous
infrared and optical data points of \citet{isr04} from 2003 October.
Filled circles are observed points, and open circles are
corrected for extinction using the X-ray
fitted $N_{\rm H}$ from each model and the \citet{pre95} relation
$N_{\rm H}/A_V = 1.8 \times 10^{21}$ cm$^{-2}$ mag$^{-1}$.
Fitting parameters of the two-component X-ray spectra are taken from Table~\ref{spectra}.
The unfolded, unabsorbed spectrum is represented here.
({\it a\/}) Power-law plus blackbody fit to the X-ray spectrum,
with $N_{\rm H} = 1.02 \times 10^{22}$ cm$^{-2}$ and $A_V = 5.7$ mag.
This panel is equivalent to Figure~2 of \citet{isr04}.
({\it b\/}) Double blackbody fit to the X-ray spectrum,
with $N_{\rm H} = 0.65 \times 10^{22}$ cm$^{-2}$ and $A_V = 3.6$ mag.
\label{broad-band}
}
\end{figure*}

According to Figure~\ref{broad-band}a, which is equivalent to  
Figure~2 of \citet{isr04}, the $K$-band measurement
and 0.5~keV flux density of \taxp\ as predicted from the power-law
fitted component can be bridged by a power law of
$F_{\nu} \propto \nu^{\alpha}$, where $\alpha = +0.6$. 
Notably, this value of $\alpha$ is in excess of the low-frequency 
limit of an optically thin
synchrotron spectrum, which has $F_{\nu} \propto \nu^{+1/3}$.
Alternatively, the standard synchrotron self-absorption mechanism,
which allows a low-frequency spectrum $F_{\nu} \propto \nu^{+5/2}$,
can be invoked to avoid exceeding the observed infrared flux.
But this would require the self-absorption frequency $\nu_a$ to fall
in the narrow (and unobservable) range $10^{16}-10^{17}$~Hz.
It also places a requirement on the radius $r$ of the
synchrotron source,
$$r = 6\times10^{14}\,B^{1/4}\,\nu_a^{-5/4}\,F_{\nu_a}^{1/2}\,d\ {\rm cm},$$
where $F_{\nu_a}$ is the flux density at $\nu_a$, extrapolated from
the fitted X-ray power law.  This relation assumes a uniform $B$
field unlike the magnetosphere of a neutron star.
But if we combine it with an
estimate of the local dipole field at radius $r$,
$$B(r) = B_s\, \left(r \over 12\ {\rm km}\right)^{-3},$$
with $B_s \approx 2\times10^{14}$~G, then the
synchrotron source must have $r < 3 \times 10^7\,d_5^{4/7}$~cm.
Such small values of $r$ are inconsistent with the synchrotron
assumption since the associated magnetic field $B(r) > 10^{10}$~G
is large enough that the soft X-ray spectrum should instead be a 
broad cyclotron feature. These serious problems with a synchrotron model
cast doubt on the utility of a steep power-law fit.

Alternatively, while \citet*{tho02} propose that multiple resonant Compton 
scattering of thermal photons can generate a high-energy tail,
this mechanism does not address the origin of the required low-energy excess,
which is just as much a part of the steep power-law component
as the high-energy tail.  Therefore, we conclude that a power law is not a 
correct description of the actual spectrum.  While it fits adequately
in the limited X-ray band, it has no compelling physical explanation
and is in conflict with lower-energy measurements.  Unfortunately,
the large column densities to most AXPs prevent an accurate
characterization of their intrinsic soft X-ray spectra
because the corrections for absorption become 
highly uncertain when they are large.

The blackbody component in the
power-law plus blackbody model is easier to accept
since its effective area is initially
$\approx 6 \times 10^{11}\,d_{5}^2$~cm$^{2}$ (neglecting
unknown geometric and beaming factors), decreasing to
$\approx 3 \times 10^{11}\,d_{5}^2$~cm$^{2}$ in the
third observation.
Because these areas are both small compared to the surface
of the NS, the large pulsed fraction of this component,
which dominates the flux from 2--5~keV and does not change as
the flux declines, can be understood as rotational modulation of a
surface hot spot.

The pulsed light curves provide an
independent diagnostic of emission mechanisms.  The strict
pulse-phase alignment and subtle change of pulse shape as a
function of energy in Figure~\ref{pulse} is unnatural under the hypothesis of
two entirely different emission mechanisms and locations.  The power-law
component of the spectral fit dominates below 1.5~keV
while the blackbody component dominates between 2 and 6 keV.  The
observed monotonic increase in pulsed fraction as a function
of energy is not accounted for in such a spectral model.
This problem applies to AXPs in general, as discussed in detail
by \"Ozel, Psaltis, \& Kaspi (2001).  A hard X-ray tail can arise from 
radiative transfer effects in a realistically modeled NS atmosphere
\citep*{oze01a,oze01b,per01}, thus reducing the need for a power-law component.
This raises the possibility that a purely thermal model may account for the
spectra and pulse profiles.

\subsection {Advantages of a Purely Thermal X-ray Spectrum}

In the alternative double blackbody spectral model, a large area
of $\approx 60\%$ the surface of the NS
is implied for the cooler blackbody component,
while the hotter component has an initial area of
$\approx 4\%$ of the NS surface. 
In its favor, a purely thermal model
can explain why the pulsed fraction increases with energy, and why the
pulses are in phase at all energies, assuming the geometry to be that
of a small hot spot surrounded by a cooler annulus.  
The phase-resolved spectroscopy discussed
in \S 2.1 supports this picture.  Also,
the extrapolated X-ray spectrum in this model does not exceed the
observed IR fluxes (Fig.~\ref{broad-band}b).  IR and optical emission from AXPs,
including \taxp, are too bright to be surface thermal emission,
but are likely to be magnetospheric synchrotron radiation.  Unlike
the X-rays, the optical/IR emission does not exceed the spin-down
luminosity, and could be rotation-powered in quiescence
\citep{oze04}, although additional energy deposited by magnetic
field decay could increase its brightness during outbursts.

Parenthetically, we mention the case of the AXP 1RXS J170849.0--400910.
A double blackbody was also one of the models
fitted to the X-ray spectrum
of 1RXS J170849.0--400910 by \citet{isr01}, with an interpretation similar 
to ours.  On the other hand, 1RXS J170849.0--400910 is unique among AXPs
for having a strong energy dependence of its pulse phase,
shifting by $\approx 0.3$ cycles
between 2 and 10~keV.  \citet{rea03} interpret this as most likely due to
a phase dependence of a power-law spectral index.  Interestingly, this
is the only AXP for which good evidence of a cyclotron absorption
feature is found, at 8.1~keV \citep{rea03}, and therefore a true
deviation from blackbody emission.

While a purely thermal model may be favored by the simple
arguments presented above,
detailed modeling of the light curves will
be needed to verify that the softest X-rays from \taxp,
which come from the cooler, larger area blackbody, can have a pulsed
fraction as high as the observed 38\%.  The emitting area might
be even larger if realistic atmosphere models are fitted instead of
pure blackbodies \citep{per01}.
One possibility is that $d$ is considerably less than 5~kpc,
thus reducing the emitting area.  Another important consideration
is the effect of realistic radiative transfer in highly magnetized
atmospheres. \citet{oze01b} showed that the sorts of pulsed light curves
that are observed from \taxp\ and other AXPs
are inconsistent with two antipodal hot spots on the surface.
However, they showed that a single hot spot
is capable of producing large pulsed fractions as well as
the observed energy dependence from a nearly orthogonal rotator
if the magnetic field is assumed to be normal to the surface.
[Note that \citet{oze01b} use a different definition of pulsed
fraction from the one that we use in Figure~\ref{pulse}].
The large predicted pulsed fractions rely on the
``beaming'' effect of anisotropic
radiative transfer in a large magnetic field, in which the
electron scattering cross sections of both the
ordinary and extraordinary modes of polarization are minimized
where the angle between the wave vector {\bf k} and the magnetic
field {\bf B} is zero.  

The quiescent \rosat\ spectrum of \taxp\
is fitted by a simple blackbody comparable to the
full surface area of a NS.
Note that the temperature measured by \rosat\
using a single blackbody model is even lower that found for the
cooler component of the two-temperature \xmm\ fit. 
This perhaps accounts for the
failure to detect rotational modulation in the quiescent state.
Thus, it is possible
that the X-ray manifestation of the outburst of a TAXP is
largely due to a sudden thermal heating event in the crust,
while the long-term quiescent flux is powered by slow dissipation
of the core magnetic field \citep{tho96} or simply by residual cooling
in the presence of an insulating atmosphere of hydrogen or helium
\citep{hey97}.

\subsection{Afterglow Following Deep Crustal Heating}

Similar to \taxp, a decay of the thermal component in the AXP 1E~2259+586
was seen during the bursting episode in 2002 June \citep{woo04},
when a small, hot region of $kT_{\rm BB} \approx 1.7$~keV and
$R \approx 1$~km cooled within one day, leaving a 
quiescent thermal component of $kT_{\rm BB} \approx 0.5$~keV
and $R \approx 6$~km.  After 1 day, the 2--10 keV 
X-ray flux decayed further as a slow $t^{-0.22}$ power law
for at least 200 days.  Hot spots
due to deep crustal heating by localized magnetic field decay,
or surface bombardment by accelerated particles, can plausibly 
produce the observed spectra and pulse profiles.  However,
heating by magnetic field decay is preferred over particles
accelerated only on open field lines
in the dipole geometry, since the canonical dipole polar cap area
is only $2 \times 10^8\,(P/5.5\,~{\rm s})^{-1}$~cm$^2$ 
for such slow rotators as these AXPs, much smaller than the detected areas.

\citet{eic03} reviewed mechanisms for ``afterglows''  from cooling
of the crust following SGR like outbursts in magnetars.
Deep crustal heating that is subject to a long conduction
time to the surface can plausibly be the cause.  \citet{tho02}
showed that the resulting cooling would follow a $t^{-0.7}$ power
law.  The initial heating could take place within
$10^4$~s due to large-scale shearing of the crust by magnetic
stresses.  \citet*{lyu02} assumed a deposition of $\sim 10^{25}$
ergs~cm$^{-3}$ to a depth of 500~m, which is $\leq 1\%$ of
the magnetic energy density at the surface.  If occurring
over the entire surface, this is $\sim 1 \times 10^{43}$
ergs, comparable to the total X-ray energy emitted during the decay 
of \taxp.  If such releases occur with recurrence time $>20$~yr,
AXPs could power transient outbursts for $>5000$~yr.
However, this simple rate estimate does not take into account
heat conduction into the core of the NS.
\citet{lyu02} find that only 20\% of the deposited
heat is radiated during the afterglow; the rest
is conducted into the interior and contributes
to the average quiescent X-ray or neutrino emission.
The observed quiescent luminosity of
$1.3 \times 10^{34}$ ergs~s$^{-1}$ could be powered by
$\sim 7 \times 10^{45}$ ergs of magnetic energy dissipated
in the core over the presumed $\sim 17,000$~yr age of the NS.

This theory has been applied to the
slow cooling observed from SGR 1900+14 \citep{ibr01,woo01,len03}
and SGR 1627--41 \citep{kou03,lyu02} following their outbursts.
In the spectra of SGR 1900+14, a blackbody component
was identified that has a high initial temperature,
$\approx 2$~keV, and radius of $\approx 1.7$~km, much
smaller than a neutron star, showing that heating (or heat
conduction or radiation) can indeed be localized.
Therefore, it seems that the long 
timescale decay of the flux from \taxp\ could be due to the same
process that occurs in the outbursts of SGRs.

While there are still too few measurements
to characterize reliably the long-term decay of
the outburst of \taxp, it does seem already to be
declining faster than the $t^{-0.7}$ power law predicted
in the crustal heating model.  We find a decay index of
1.03--1.26 in the interval 9--15 months after the presumed
outburst, but an exponential of time constant 300 days
is a better description of the entire light curve
since outburst.  Similar deviations of the 
post-outburst decay curve of SGR 1627--41 from a power law
\citep{kou03} complicate the crustal heating interpretation.
Of course, the existing cooling models are based on solving
a one-dimensional heat transfer equation, and do not involve
localized heating and possible changes in emitting area such
as are suggested by the observations of \taxp.

Another variable AXP for which detailed spectral analysis at
different intensity levels has been performed is 1E~1048.1--5937.
\citet{mer04} found that its spectral shape remained invariant
between states of luminosity that differed by a factor of 5.
In a fitted two-component model, the power-law index as well
as the blackbody temperature remained the same, which
implies that it is not realistic to assume two physically distinct
emission mechanisms.  Instead, the variability in flux could be explained
in a purely thermal model by a change in emitting area.
Such an interpretation would be consistent with the pulsed fraction of
the light curve of 1E~1048.1--5937, which decreased as the flux increased
\citep{mer04}.  We interpret the decay of \taxp\ along the same lines.
 
\section{Conclusions and Future Work}

We used three \xmm\ observations of \taxp\ obtained during its decline from
outburst to make a number of quantitative measurements, although their
robustness needs to be tested with additional follow-up observations.
The decay is consistent with an exponential of time constant
$\approx 300$ days.  The pulsar spin-down rate is quite variable, having
decreased to $(5.1 \pm 1.6)\times10^{-12}$ s~s$^{-1}$ after initial measurements
by \xte\ found it to vary in the range $(1-2) \times10^{-11}$ s~s$^{-1}$.
The inferred characteristic age $\tau_{\rm c} \approx 17,000$~yr,
magnetic field strength $B_{\rm s} \approx 1.7\times 10^{14}$~G,
X-ray spectra, and pulsed light curves, are all consistent with
the behavior of an anomalous X-ray pulsar, albeit one that has varied
in flux by two orders of magnitude.  Assuming that the decay continues
at the observed exponential rate, the total energy radiated in the present
outburst will be $\approx 1.5 \times 10^{43}\,d_5^2$~ergs.  This is
comparable to the heat that can be deposited in the crust
by conversion of $<1\%$ of the magnetic energy density at the surface.
Such a magnetar mechanism could power outbursts similar to the
current one at $>20$~yr intervals for a large fraction of the
characteristic age of the pulsar, while the quiescent luminosity observed
in a number of archival X-ray observations could come from the long-term
average magnetic decay heating of the NS core.  Models of deep crustal
heating typically predict $t^{-0.7}$ power-law decays,
which are not matched by the sparse measurements so far
obtained from \taxp.  However, more complex decay behavior has been
seen in several other magnetars, which may require refinements
to the initial simple models, in particular, generalization to
two dimensions.

The X-ray spectrum and pulse profiles of \taxp\ are typical of
AXPs.  We discussed the problems with allowing a power-law plus
blackbody fit, which results in a steep power law that is difficult
to reconcile with the faintness of the optical/IR counterpart.  The lack
of variation in pulse profile as a function of energy also
argues against the action of two different emission mechanisms.
Instead, a purely thermal model for the spectrum, here adequately
modeled as two temperatures, is more physically motivated.
The required blackbody areas are less than or equal to
the surface area of a neutron star, and the pulse modulation
can probably be understood in terms of anisotropic heat conduction
and radiative transfer in the strong magnetic field.  Two dimensional
cooling models will be required to explain why the decay is characterized
as a decline in emitting area rather than temperature, the latter remaining
constant, at least initially.  The spectral
data obtained so far are limited in that the existing \xmm\ observations
span slightly less than a factor of 2 in luminosity.  The most stringent
constraints on models may yet come from X-ray spectra and pulse profiles 
to be obtained over the coming years, as the luminosity of \taxp\
is still a factor of $20-30$ above its historic quiescent level.

The existence of serendipitous archival data allowed us to determine that
\taxp\ lacked pulsations in the quiescent state preceding the current
outburst.  This suggests that CCOs may be interpreted as quiescent AXPs.
No periodic signals have been
detected so far from the CCOs in the Cas~A, Puppis~A, 
and Kes~79 SNRs, for example.  Cas~A is known to
be a very young, $\approx 300$~yr object; its central source
has a spectrum consistent with an AXP \citep*{cha01,mer02}.
A similar object may be the point source at the center of
Kes~79 \citep{sew03}.
The radio quiet neutron star RX~J0822--4300 in Puppis A also
has a spectrum similar to that of \taxp\ \citep{bec02}.
\citet*{gae00} argue that RX~J0822--4300 may have AXP-like
spin parameters.  
The CCO in SNR RCW~103, a non-pulsating AXP-like object, has been
monitored for over a decade and it is found to display marked 
variability on months to hours time scales.
[While a possible 6~hr periodic variation suggests that this object
is a low-mass binary \citep{bec02,san02}, the evidence is not
conclusive.]
SGRs are hypothesized to be older ($\sim 10^4$ yr) manifestations of
the AXPs \citep*{got99b,gae01}. AXPs and SGRs have long been
considered related phenomena, reinforced by the recent detection of
SGR-like bursts from the two AXPs 1E~1048.1--5937 and 1E~2259+586
\citep*{gav02,kas03}.  The various ``classes'' of young NSs that
differ significantly from rotation-powered pulsars
are phenomenologically related, possibly through an evolutionary
progression.

Although TAXPs are relatively rare, their short active
duty cycle suggests the existence of a
larger population of unrecognized young NSs.  To this end,
\taxp\ provides a unique window into this population, with
prior measurements in the quiescent state and detailed observations
during its active, pulsed phase.  It is important to monitor closely
the fading of \taxp\ during this possibly one-time opportunity to
study the processes of thermal and non-thermal emission
to which magnetars convert their energy.
Measurement of the quiescent spectrum is especially
important to help identify the faint population of likely missed NSs.

\acknowledgements

This investigation is based on observations obtained with \xmm,
an ESA science mission with instruments and
contributions directly funded by ESA Member
States and NASA.  We thank R. Warwick and D. J. Helfand for providing
the 2003 October data from the \xmm\ Galactic Plane Survey.
This work was also supported by 
NASA \xte\ grant NNG04GF59G to J.P.H., and
NASA LTSA grant NAG~5-8063 to E.V.G.

\end{document}